\documentclass{article}

\usepackage{color}
\usepackage{amsmath}
\usepackage{graphicx}
\usepackage{url}

\date{}

\newcommand{\alina}[1]{{ \color{black}  #1}}

\begin{document}
\vspace*{0.2in}

\begin{flushleft}
{\Large
\textbf\newline{Algorithmic bias amplifies opinion polarization: A bounded confidence model} 
}
\newline
\\
Alina S\^irbu\textsuperscript{1,2},
Dino Pedreschi\textsuperscript{1},
Fosca Giannotti\textsuperscript{3},
J{\'a}nos Kert{\'e}sz\textsuperscript{4,5},
\\
\bigskip
\textbf{1} Department of Computer Science, University of Pisa, Pisa, Italy
\\
\textbf{2} Science Division, New York University Abu Dhabi, Abu Dhabi, United Arab Emirates
\\
\textbf{3} Istituto di Scienza e Tecnologie dell'Informazione ``A. Faedo" - CNR, Pisa, Italy
\\
\textbf{4} Center for Network Science, Central European University, Budapest H-1051, Hungary
\\
\textbf{5} Department of Theoretical Physics, Budapest University of Technology and Economics, Budapest H-1111, Hungary
\bigskip

* alina.sirbu@unipi.it

\end{flushleft}

\section*{Abstract}
The flow of information reaching us via the online media platforms is optimized not by the information content or relevance but by popularity and proximity to the target. This is typically performed in order to maximise platform usage. As a side effect, this introduces an algorithmic bias that is believed to enhance polarization of the societal debate. To study this phenomenon, we modify the well-known continuous opinion dynamics model of bounded confidence in order to account for the algorithmic bias and investigate its consequences. In the simplest version of the original model the pairs of discussion participants are chosen at random and their opinions get closer to each other if they are within a fixed tolerance level. We modify the selection rule of the discussion partners: there is an enhanced probability to choose individuals whose opinions are already close to each other, thus mimicking the behavior of online media which suggest interaction with similar peers. As a result we observe: a) an increased tendency towards polarization, which emerges also in conditions where the original model would predict convergence, and b) a dramatic slowing down of the speed at which the convergence at the asymptotic state is reached, which makes the system highly unstable. Polarization is augmented by a fragmented initial population.

\section{Introduction}
Political polarization is a generally observed, ingravescent negative trend in modern western societies \cite{PewReport2014, Quartz2016, Andris2015,Delvicario2017} with such concomitants as ``alternative realities", ``filter bubbles", ``echo chambers", and ``fake news". Several causes have been identified (see, e.g., \cite{Campbell2016}) but there is increasing evidence that new online media is one of them \cite{Prior2007,Baum2008,Graber2012}. Earlier it was assumed that traditional mass media mostly influence the politically active elite of the society and they only indirectly affect polarization of the entire population. The recent dramatic changes with the occurrence of online media, the ubiquity of the Internet with all the information within reach of a few clicks and the general usage of online social networks have increased the number of communication channels by which political information can reach citizens. Somewhat counterintuitively, this has not lead to a more balanced information acquisition and a stronger tendency towards consensus, as argued in \cite{Messing2012}, on the contrary. One of the reasons may be that the new media have enhanced the reachability of people, which can be used for transmitting simplified political answers to complex questions and thus act toward polarization \cite{Charles2012}. Moreover, the stream of news is not organized in the new media in a balanced way, but by algorithms, which are built to maximise platform usage. It is conjectured that this generates an ``algorithmic bias", which artificially enhances opinion polarization. This is an artefact of online platforms, also called "algorithmic segregation" \cite{Ignatieff2016}. The link between opinion polarization and algorithmic bias from online platforms has not been proven to date, so the aim of the present paper is to study this effect by a simple, bounded-confidence-type opinion dynamics model.

A considerable and rapidly increasing part of the population does not use traditional media (printed press, radio, TV or even online journals) for obtaining news \cite{Purcell2010,Hermida2012,Villi2015} but turns to the new media like online social networks or blogs. However, the flow of news in the new media is not selected by the information value but rather by popularity, by "likes" \cite{Pariser2011}. As people tend to identify themselves with views similar to their own and will like corresponding news with higher probability, it is in the interest of the service providers to channel the information already in a targeted way. This means users do not even get confronted with narratives different from their favorite ones. Large effort is paid to develop efficient algorithms for the appropriate channeling of the news to provide the stream that has the highest chance to collect maximum number of likes.

Another important factor pointing in the same direction is related to the "share" function, which is largely responsible for the fast spreading of news and thus enhancing popularity. This spreading takes place on the social network, where links are formed mostly as a consequence of homophily, i.e., exchange of information takes place between people with similar views. The variety of interactions of a person (family, school, work, hobby, etc.) may contribute to diversification of information sources \cite{Bakshy2015}, nevertheless, regarding political views homophily seems particularly strong \cite{Colleoni2014}.
 The sharing among users with similar beliefs was also shown to be true in the context of spreading of misinformation on Facebook, leading to an echo-chamber effect~\cite{Vicario2016}, \alina{as well as on Twitter where partisan users were demonstrated to have an important role in polarization~\cite{Garimella2018}.}

Network effects are obviously very important in the spreading of news and opinions, however, in our present study we will ignore the network structure of the system and will exclusively focus on the consequences of the algorithmic bias for selecting the content presented to the users. This corresponds to a mean field approach, which is widely used as a first approximation to spreading problems \cite{Pastor-Satorras2015}.

The task is therefore to model the evolution of the attitude of the society, provided there is a bias in selecting partners whose opinions are confronted. 
Recent years has seen the introduction of several models of opinion dynamics~\cite{Sirbu2017}, however none of them includes algorithmic bias in the sense described here. Hence we provide an extension of one of the existing models to study this effect.
For the sake of simplicity we use a bounded confidence opinion dynamics model \cite{Deffuant2000}, which is known to be able to describe both consensus and polarization of opinions depending on the tolerance level of the agents. In the mean field version two agents are selected at random and their opinions, represented by real numbers, get closer if they are within the tolerance level. The bias is introduced such that already at the selection of the partners their opinions are considered and agents with closer opinions are selected with higher probability. With this simple modification we see two effects. First, the tendency toward consensus is hindered, i.e., larger tolerance level is needed to achieve consensus; second, the approach to the asymptotic state is slowed down tremendously. Therefore, the selection bias affects the opinion formation process in two profound ways: a) by exacerbating polarization, which emerges also in conditions where the original model would predict consensus, and b) by slowing down the spreading process, which makes the system highly unstable.

The paper is organized as follows. In the next section we introduce the model in detail. Section Results contains the results of the simulations. We close the paper with a discussion and an account of further research.

\section{The model}

The original bounded confidence model~\cite{Deffuant2000} considers a population of $N$ individuals, where each individual $i$ holds a continuous opinion  $x_i \in [0,1]$. This opinion can be considered the degree by which an individual agrees or not to a certain position. Individuals are connected by a complete social network, and interact pairwise at discrete time steps. The interacting pair $(i,j)$ is selected \emph{randomly} from the population at each time point $t$. After interaction, the two opinions, $x_i$ and $x_j$ may change, depending on a so called \emph{bounded confidence parameter} $\varepsilon \in [0,1]$. This can be seen as a measure of the open-mindedness of individuals in a population. It defines a threshold on the distance between the opinion of the two individuals, beyond which communication between individuals is not possible due to conflicting views.

If we define the distance between two opinions $x_i$ and $x_j$ as 
$ d_{ij}=|x_i-x_j| $, then information is exchanged between the two individuals only if $d_{ij} \leq \varepsilon$, otherwise nothing happens. The exchange of information results in the two opinions becoming closer to one another, modulated by a convergence parameter $\mu \in (0,0.5]$

\begin{equation*}
x_i(t+1)=x_i(t)+\mu(x_j(t)-x_i(t))  
\end{equation*}
\begin{equation*}
x_j(t+1)=x_j(t)+\mu(x_i(t)-x_j(t)) 
\end{equation*}
In the following we consider only the case of $\mu=0.5$, hence when both individuals take the average opinion.

To introduce algorithmic bias in the interaction between individuals, we modify the procedure by which the pair $i,j$ is selected. In the original model, $i$ and $j$ are selected uniformly at random from the population. Instead, algorithmic bias makes encounters of similar individuals more probable. To account for this, we first select  $i$, then the selection of $j$ is performed with a probability that depends on $d_{ij}$:
\begin{equation}\label{eq_prob}
p_i(j)=\frac{d_{ij}^{-\gamma}}{\sum_{k \neq i}{d_{ik}^{-\gamma}}}
\end{equation}
In this way, the probability to select $j$ once $i$ was selected is larger if $d_{ij}$ is smaller, i.e. alike individuals interact more. The parameter $\gamma$ is the strength of the algorithmic bias: the larger $\gamma$, the more rare will be the encounters among individuals with distant opinions. For $\gamma=0$ we obtain the original bounded confidence model, $p_i(j)=\frac{1}{N-1}$.

In the following we will analyse the new model numerically, through simulation of the opinion formation process. To avoid undefined operations in Eq~\ref{eq_prob}, when $d_{ik}=0$ ($k \neq i$), we use a lower bound for  $d_{ik}$, $d_\varepsilon$. So, if  $d_{ik} <d_\varepsilon$ then $d_{ik}$ is replaced with $d_\varepsilon$ in Eq~\ref{eq_prob}. We use $d_\varepsilon=0.0001$ in the following.

\section{Results}

In order to understand how the introduction of the algorithmic bias affects model performance, we study the model under multiple criteria for various combinations of parameters $\varepsilon$ and $\gamma$. We are interested in whether the population converges to consensus or to multiple opinion clusters (Section~\ref{sec:clusters}), and how fast convergence appears (Section~\ref{sec:time}). For this we concentrate on the transition between one and two clusters. We also consider the influence of the size of the population on the behavior observed, both for the original and extended model (Section~\ref{sec:size}). Furthermore, the effect of a segregated initial population is studied (Section~\ref{sec:init}). For each analysis we repeat simulations multiple times to account for the stochastic nature of the model, and show average values obtained for each criterion above.

\subsection{Consensus versus opinion segregation}\label{sec:clusters}

\begin{figure}
\centering
\includegraphics[width=0.8\textwidth]{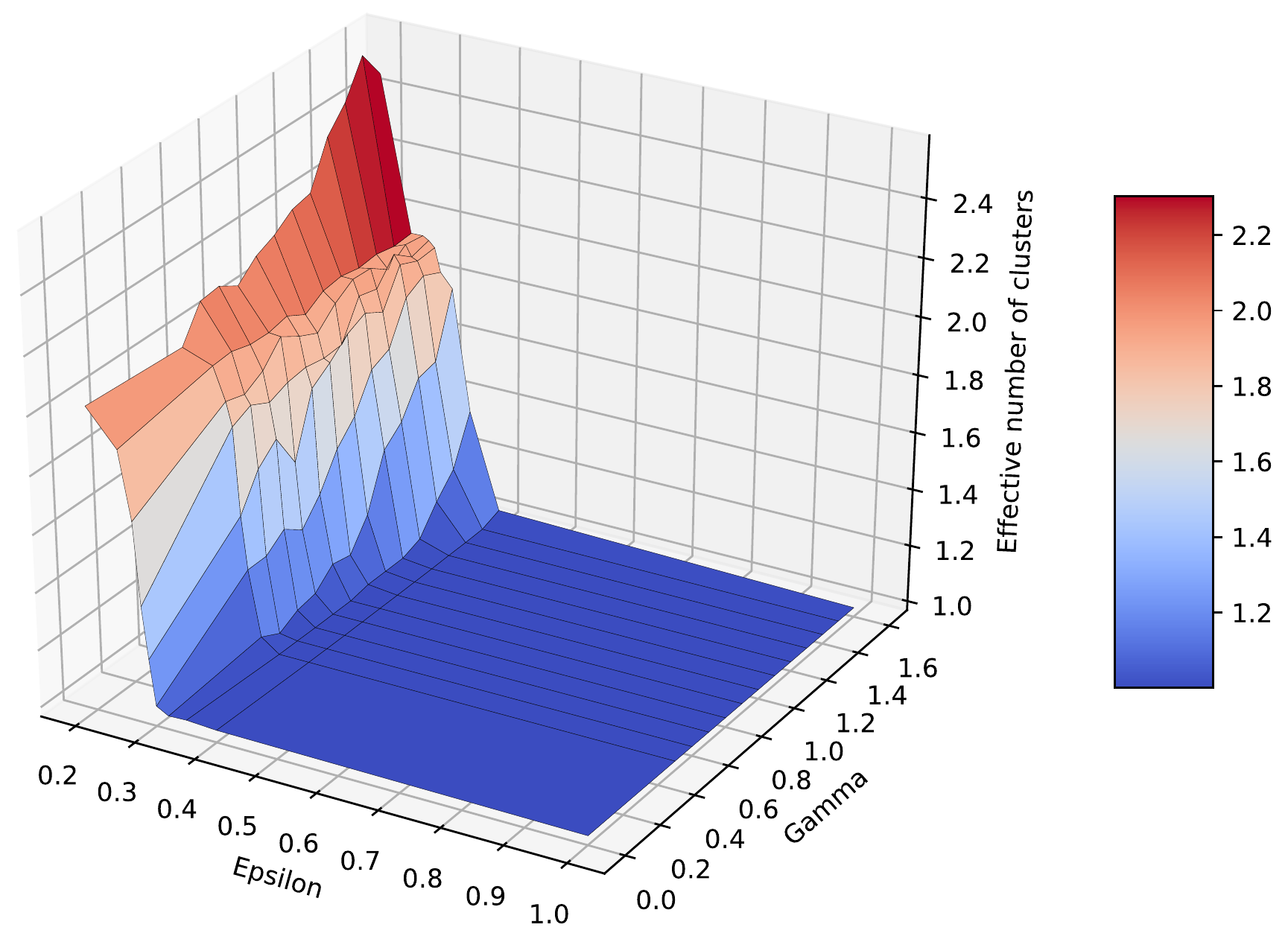}
\includegraphics[width=0.8\textwidth]{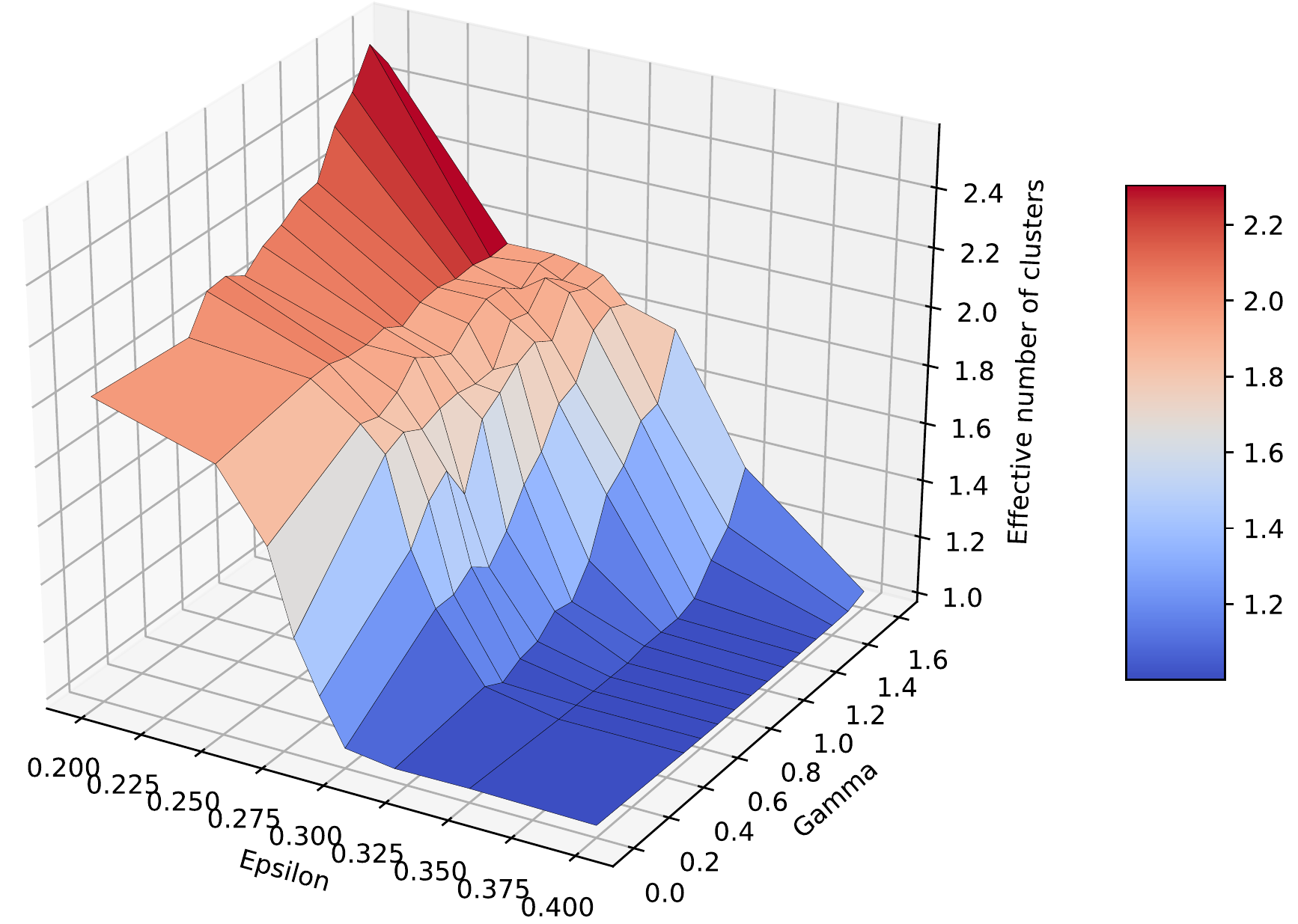}
\caption{{\bf Number of clusters obtained for various $\varepsilon$ and $\gamma$.} The top panel shows the space for $0.2\leq\varepsilon \leq1$ while the bottom panel zooms into the area where $0.2\leq\varepsilon \leq0.4$.  Values are averaged over 85 runs.  }
\label{fig_3dspace}
\end{figure}

The behaviour of the original bounded confidence model is defined by the parameter $\varepsilon$~\cite{Deffuant2000}. When this is large enough, an initially uniformly random population converges to consensus, while as $\varepsilon$ decreases, clusters emerge in the population. It was shown that the number of major clusters can be approximated by $\lfloor \frac{1}{2\varepsilon} \rfloor$.  This approximation ignores minor clusters that may emerge in some situations \cite{Lorenz2007continuous}.

In the following we analyse the number of clusters obtained for our extended model, starting from an uniform initial distribution of opinions, for a population of size $N=500$. We concentrate on the area in the  $(\varepsilon, \gamma)$ space where, in the original model, the number of clusters is smaller or equal to 2, i.e. $\varepsilon \geq 0.2$. In order to quantify the number of clusters, given the existence of major and minor clusters, we use the \emph{cluster participation ratio} as a criterion. This takes into account not only the number of clusters, but also the fraction of the population in each, measuring thus the \emph{effective number of clusters}. Hence two perfectly equal clusters will result in a cluster participation ratio of $2$, however if one cluster is larger, the measure will take a value in $(1,2)$. The effective number of clusters measured in this section is thus computed as:

\begin{equation}
C=\frac{(\sum_i c_i)^2}{\sum_i c_i^2}
\end{equation}  
where $c_i$ is the size of cluster $i$.

Fig~\ref{fig_3dspace} displays the effective number of clusters for various $\varepsilon$ and $\gamma$ values, using averages over multiple runs. The top panel shows results for $\varepsilon$ between $0.2$ and $1$, while the bottom panel zooms in to the area where  $\varepsilon$ is between $0.2$ and $0.4$. Note that in these simulations, the population converges to either one, two or three major clusters of similar mass, plus some minor negligible clusters. For the same parameter setting, the population may converge to one cluster in some simulations, or to two clusters in others. We present the average values as obtained for 85 independent runs. 

\begin{figure}
\centering
\includegraphics[width=0.7\columnwidth]{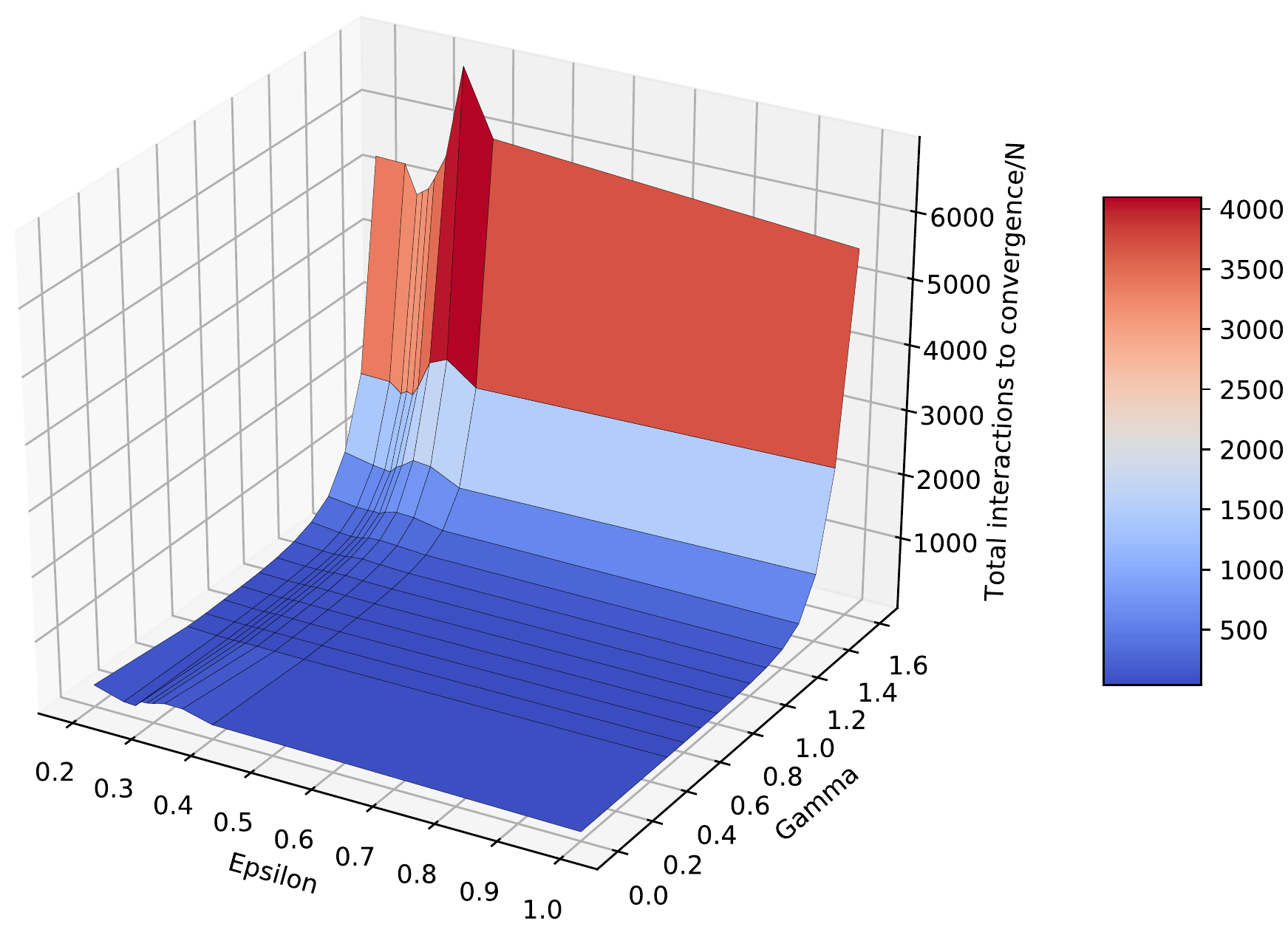}
\caption{Total number of interactions required for convergence normalized by the number of individuals, averaged over 85 runs.}
\label{fig_time3d}
\end{figure}

The plot shows that our simulations reproduce perfectly the behaviour of the original model ($\gamma=0$), where a transition between 2 and 1 cluster takes place 
for $\varepsilon \in [0.25,0.3]$. It is 
clear that the introduction of 
$\gamma>0$ causes an increase in the effective number of clusters, compared to the original model. For instance, for $\varepsilon=0.35$, the original model results in one cluster, while for our model,  new clusters start to emerge for $\gamma \geq 1.3 $. For $\varepsilon=0.32$, the transition starts even earlier before $\gamma = 1$, with the average number of clusters very close to 2 at $\gamma = 1.6$. For the case of $\varepsilon=0.2$, when the original number of clusters was already 2, $\gamma$ increases 
$C$ towards 3 opinion clusters. Hence, it appears from our simulations that algorithmic bias causes segregation in the bounded confidence model, by increasing the number of clusters 
with increasing bias.

\subsection{Time to convergence}\label{sec:time}

While the asymptotic number of opinion clusters is very important, the time to obtain these clusters is equally so. In a real setting, available time is finite, and so if consensus forms only 
after a very long period of time, it may never actually emerge in the real population. Thus, we measure the time needed for convergence 
(to either one or more opinion clusters) 
in our extended model. This can be counted as total number of pairwise interactions required to obtain a stable configuration, divided by $N$, the population size.

Fig~\ref{fig_time3d} shows how the total number of interactions required depends on both $\varepsilon$ and $\gamma$. It is clear that the time to convergence grows very fast with $\gamma$, for all $\varepsilon$ values, including the situation when the population converges to consensus. This indicates that, in a real setting with finite observation time this consensus may 
not emerge at all. Hence $\gamma$ has a double segregation effect: not only the number of clusters grows, but consensus becomes extremely slow as well.

\begin{figure}
\centering
\includegraphics[width=0.24\textwidth]{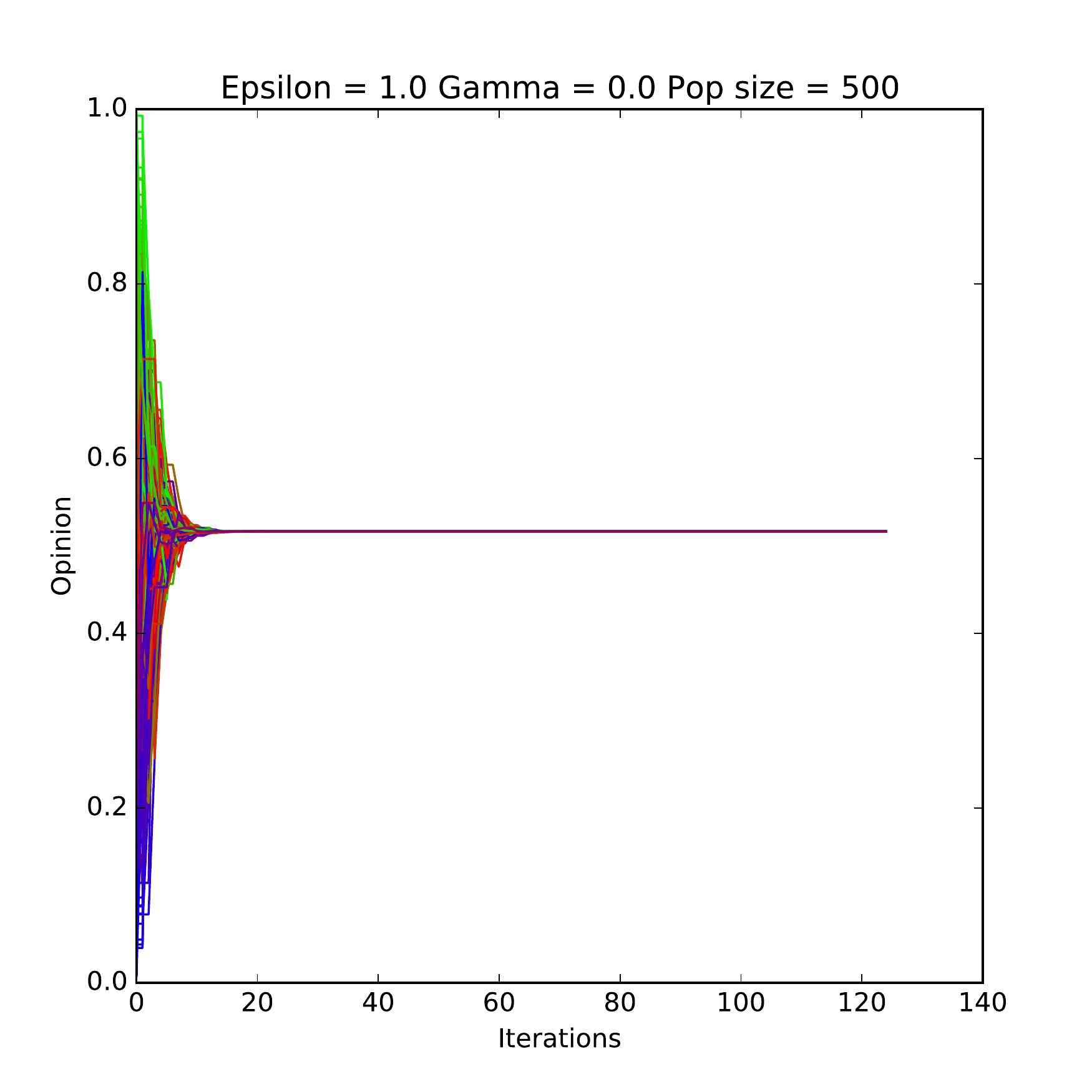}
\includegraphics[width=0.24\textwidth]{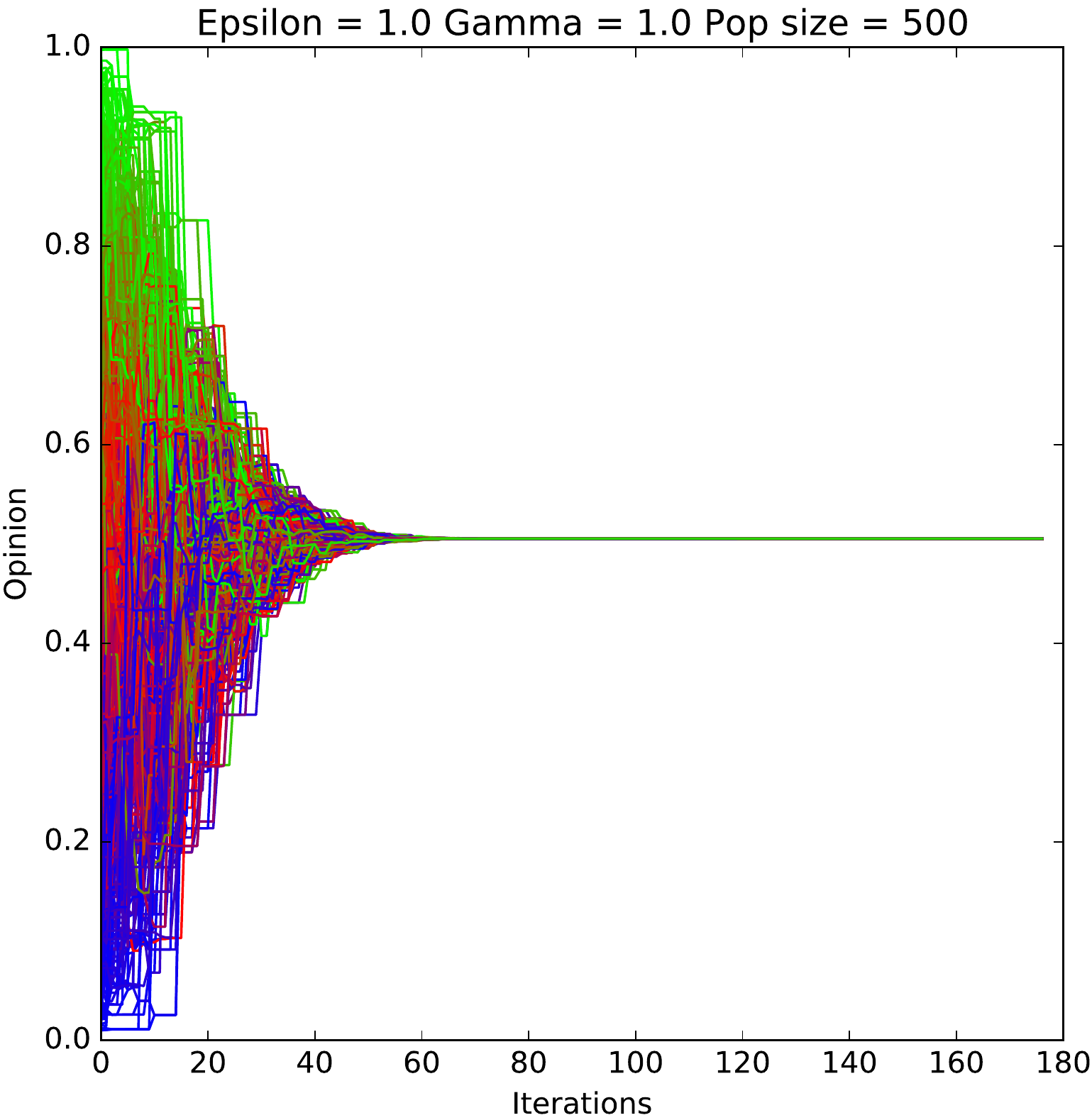}
\includegraphics[width=0.24\textwidth]{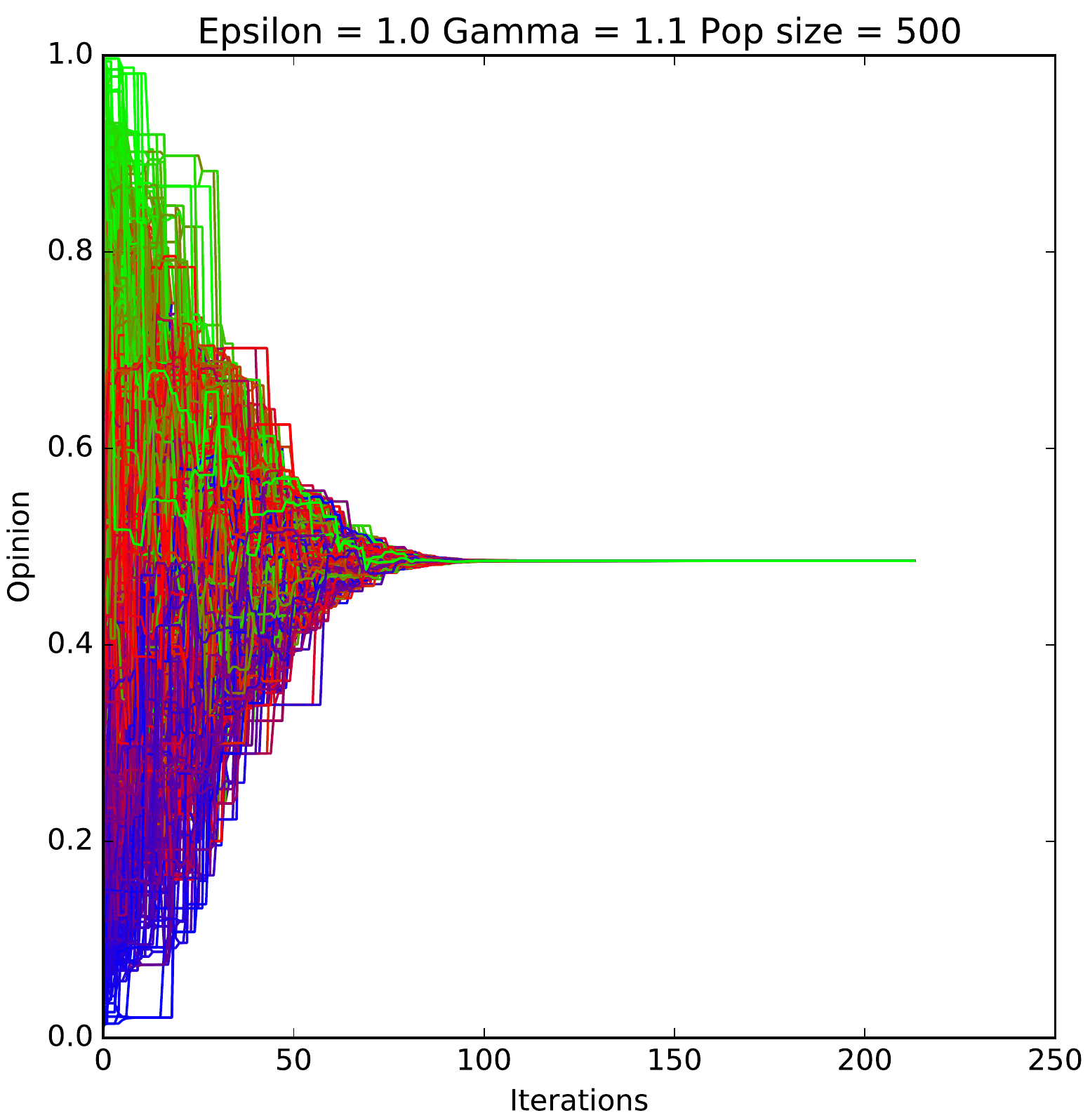}
\includegraphics[width=0.24\textwidth]{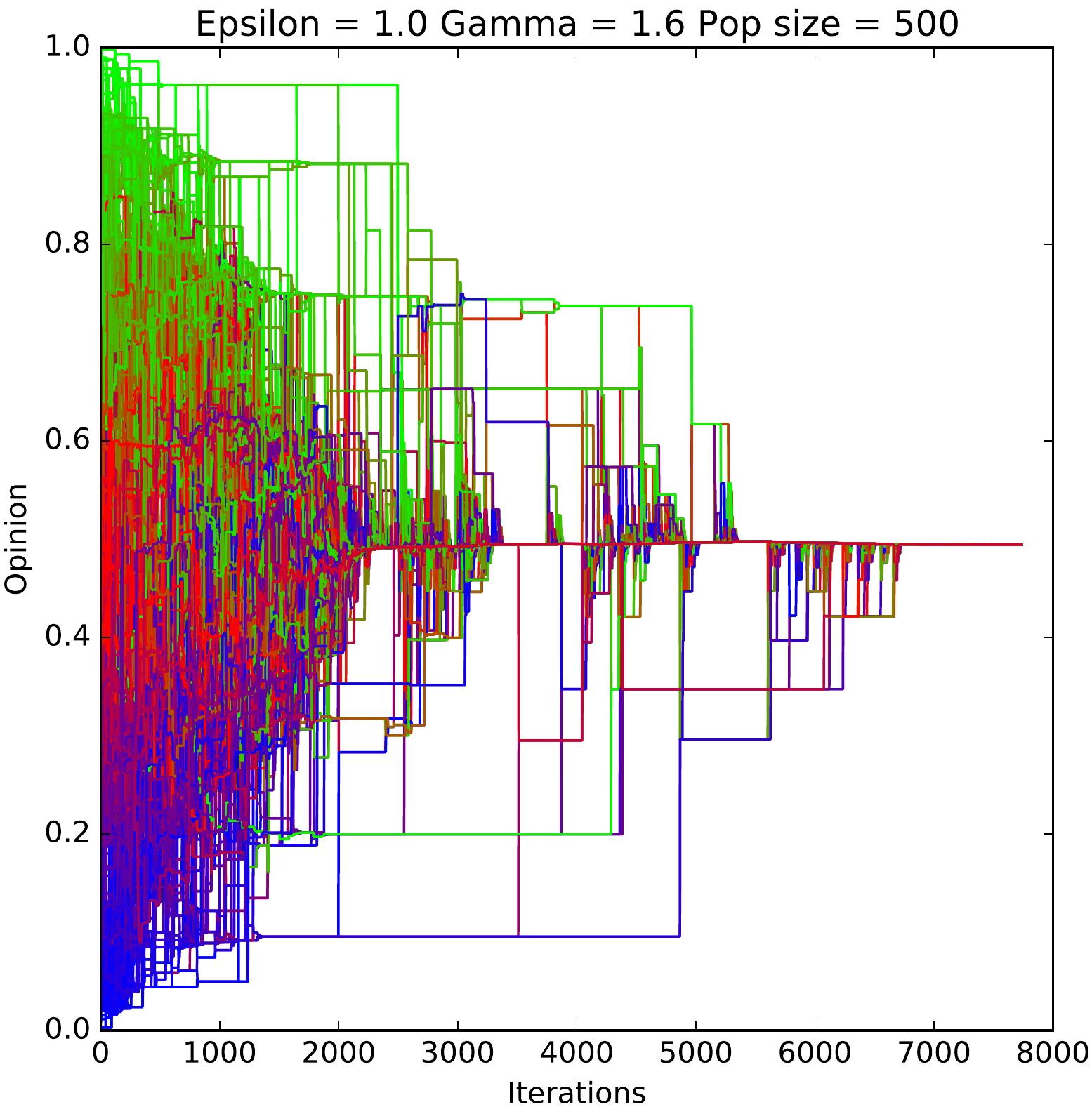}\\
\includegraphics[width=0.24\textwidth]{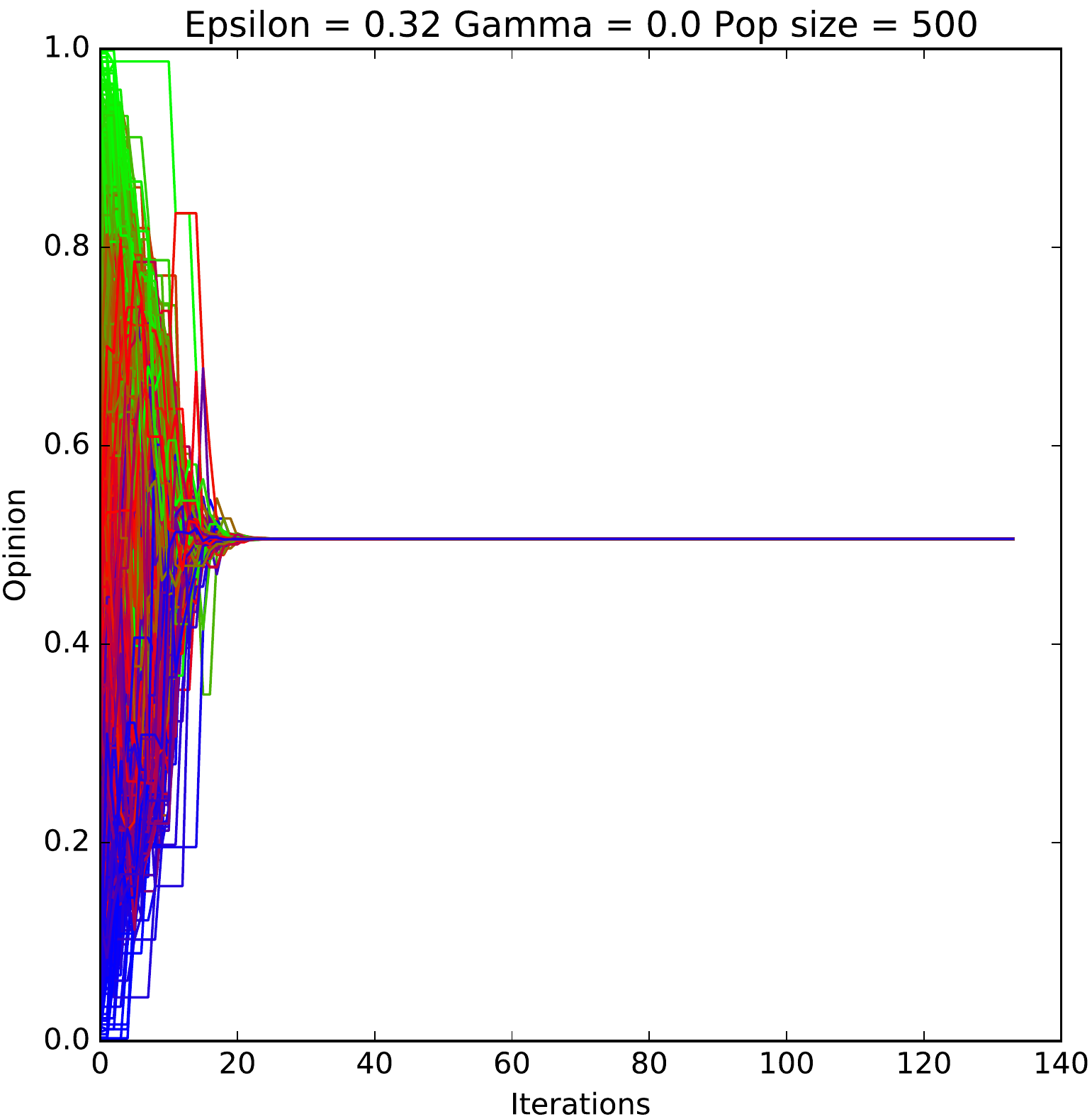}
\includegraphics[width=0.24\textwidth]{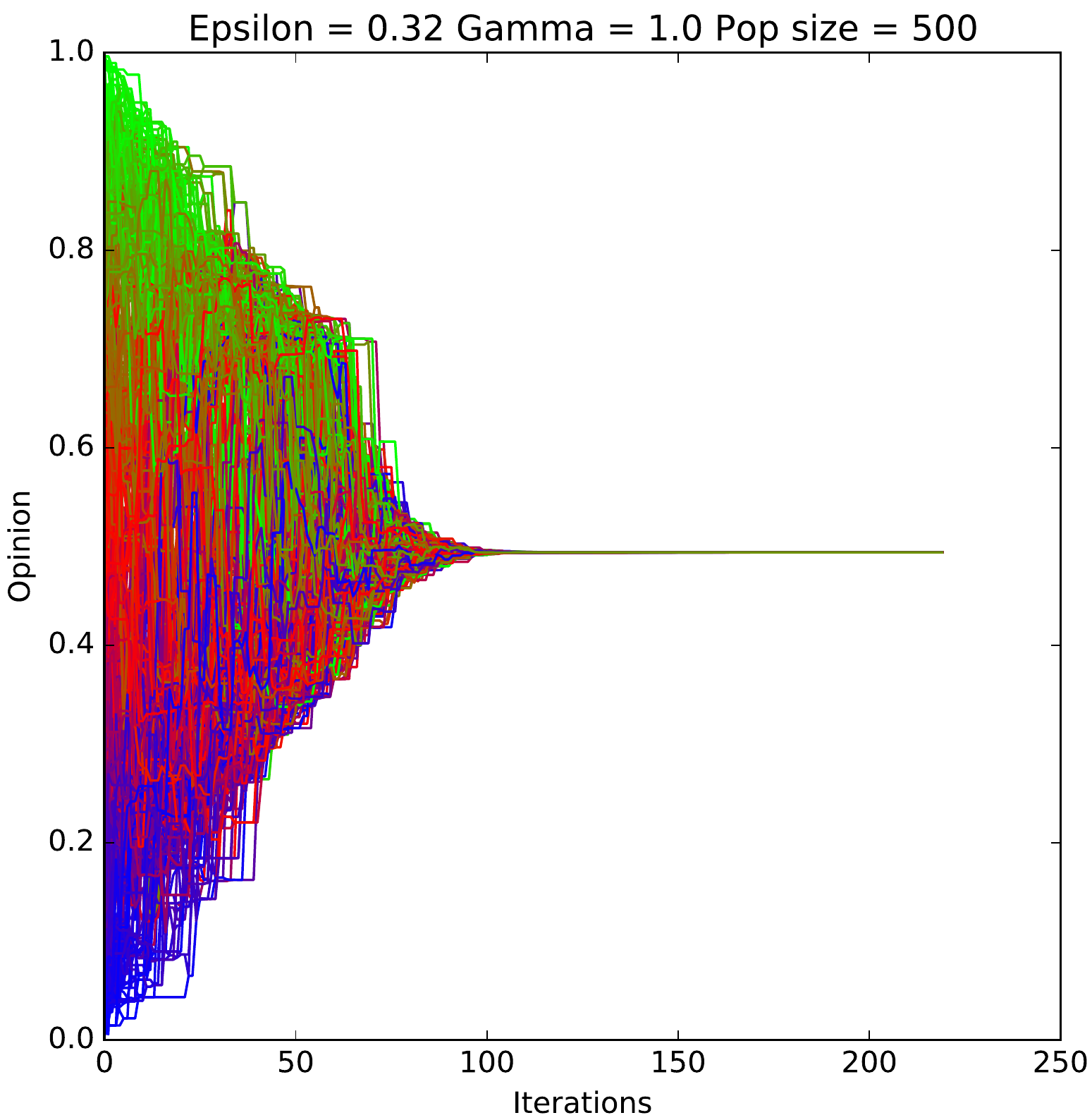}
\includegraphics[width=0.24\textwidth]{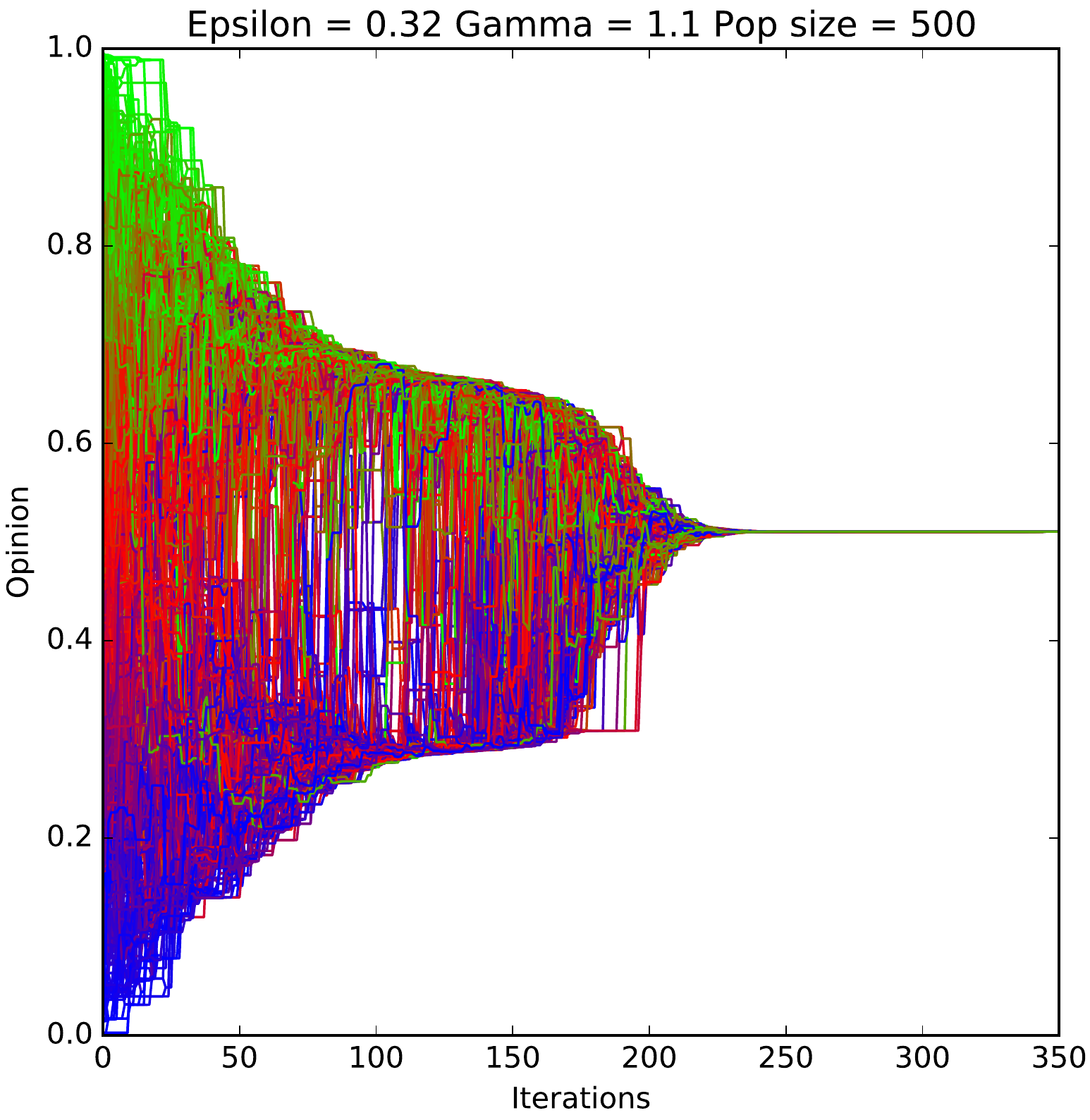}
\includegraphics[width=0.24\textwidth]{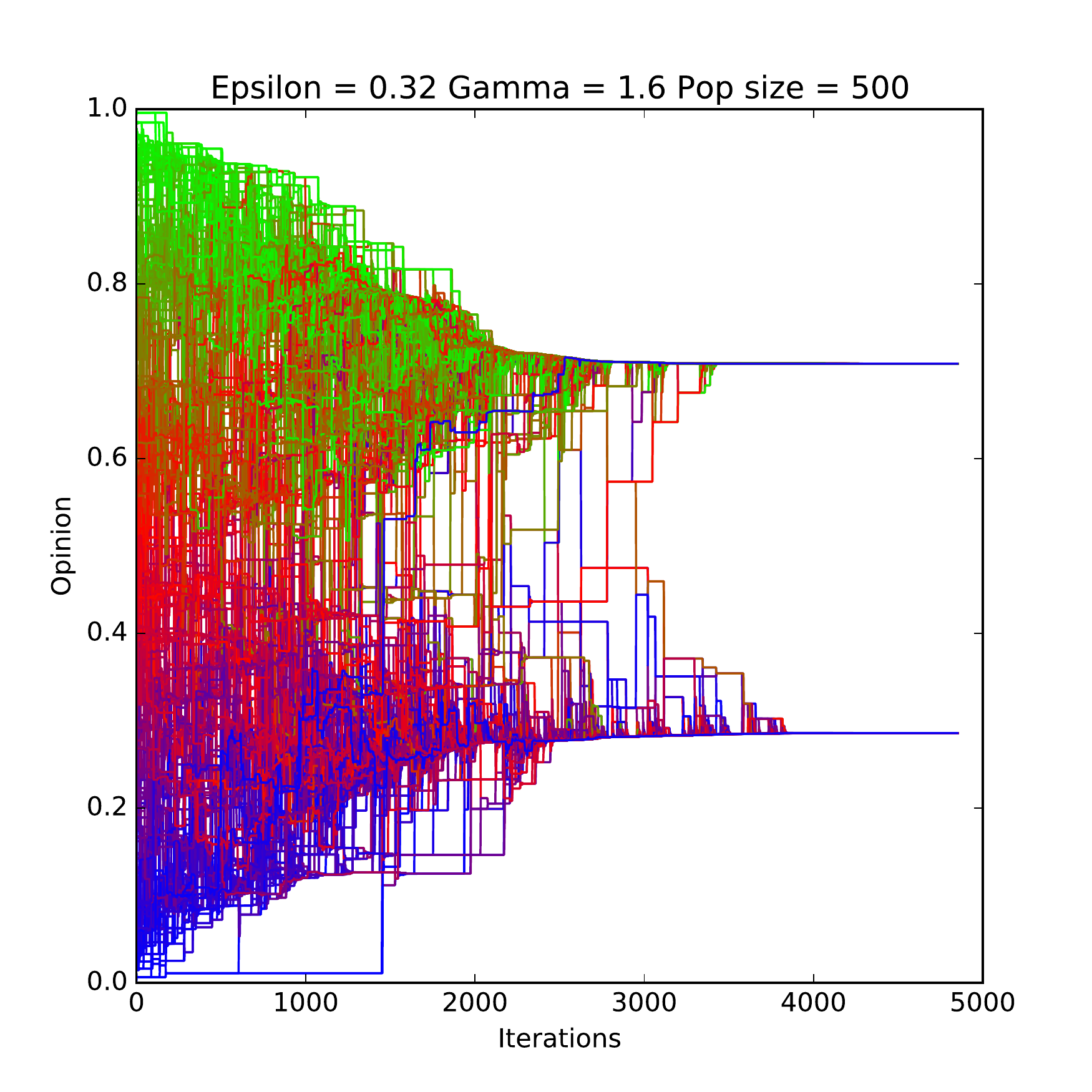}
\caption{{\bf Evolution of the population of opinions for various $\gamma$  and $\varepsilon$ values.} The first row corresponds to the case where $\varepsilon=1$, while the second row corresponds to $\varepsilon=0.32$. In both cases $\gamma \in \{0,1,1.1,1.6\}$ (left to right).
}
\label{fig_example}
\end{figure}

\alina{To support this observation, we show in Fig~\ref{fig_example} the evolution of the population for various values of $\gamma$, when $\varepsilon=1$ and $\varepsilon=0.32$. In the first case ($\varepsilon=1$), the population always converges to one cluster, however we can notice the increase in the number of iterations required, with many clusters coexisting for a long period of time before convergence. When $\varepsilon=0.32$, the original Deffuant model results in one cluster, with fast convergence. As $\gamma$ grows, convergence slows down first, and when $\gamma$ reaches a certain threshold two clusters emerge. The case where $\gamma=1.1$ is close to the transition. It is particularly interesting, because we can notice that initially two clusters coexist for a while, but they eventually merge into one cluster. In other simulation instances with the same parameter values, the two clusters never merge. }

\alina{Another observation emerging from Fig~\ref{fig_time3d} is that, besides the general fast growth of the time to convergence, we also observe a smaller peak in the convergence time for an $\varepsilon$ between 0.25 and 0.35, for all values of $\gamma$. This corresponds to a slowing down of convergence around the phase transition (from one to two clusters), which is a known physical phenomenon.}

One may argue, however, that measuring the time as the total number of interactions may inflate the figures.  Each interaction can have 3 outcomes: (1) nothing happens because a pair of individuals with identical opinions ($x_i=x_j$) were selected, (2) nothing happens because of bounded confidence $d_{ij}>\varepsilon$ or (3) the two opinions actually change.  In the following we denominate the third type of interaction as `active' interactions.

\begin{figure}
\centering
\includegraphics[width=0.7\columnwidth]{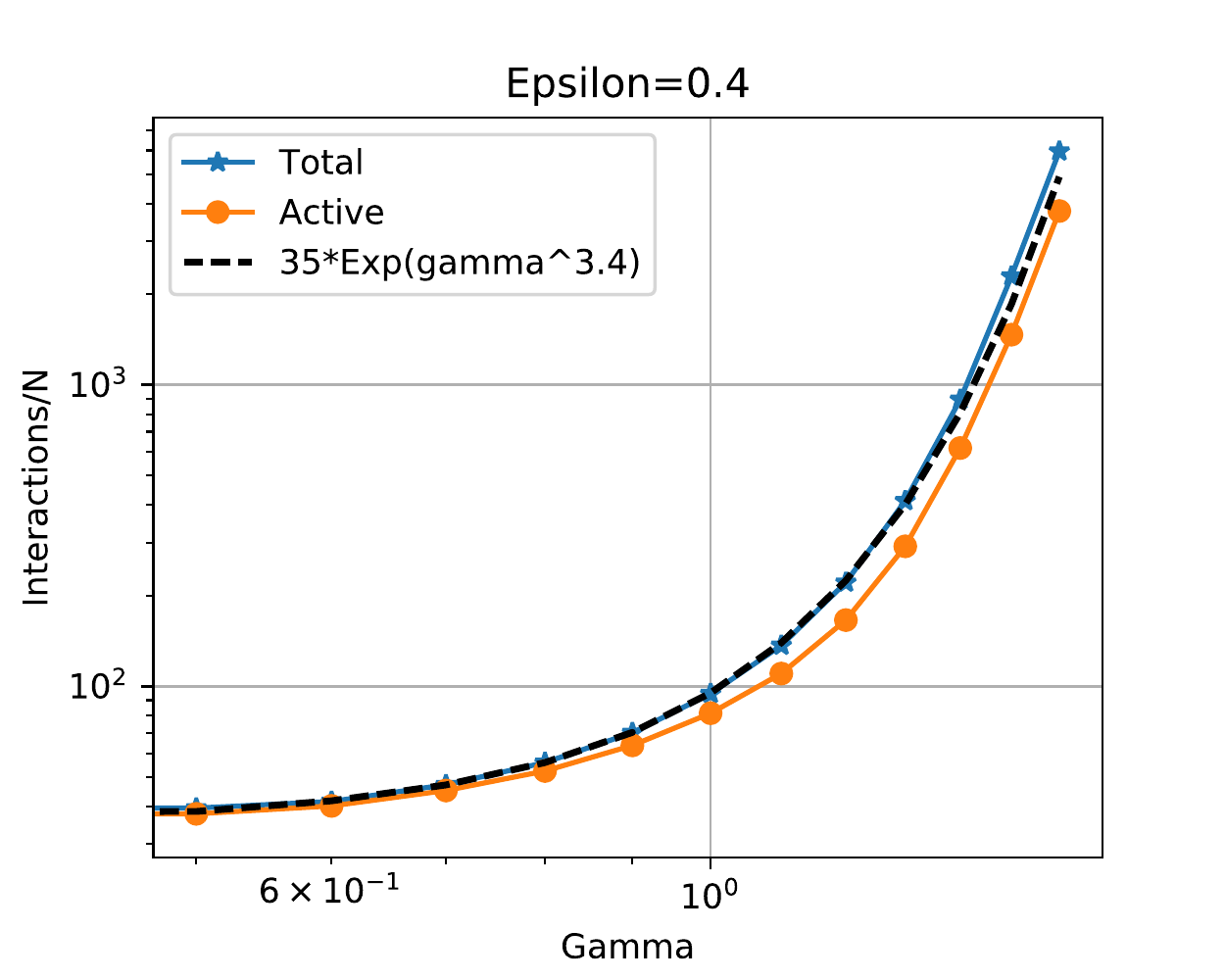}
\caption{{\bf Time to convergence. } Normalized total, non-null difference and active number of interactions required for convergence for $\varepsilon=0.4$. The reference $35\exp(\gamma^{3.4})$ is shown as a visual aid only.
}
\label{fig_time0.4}
\end{figure}

Fig~\ref{fig_time0.4} details the total and active number of interactions for the example case of $\varepsilon=0.4$. Both measures grow like $\exp(\gamma^\sigma)$ where $\sigma \sim 3.4$, with a small difference visible between active and total interactions. This extremely fast growth of the convergence time means that, in practice, consensus is hindered even by weak algorithmic bias, since consensus is slow to form, hence the population stays in a disordered state for a long time.

\subsection{Finite size effects}\label{sec:size}

\begin{figure}
\centering
\includegraphics[width=0.7\columnwidth]{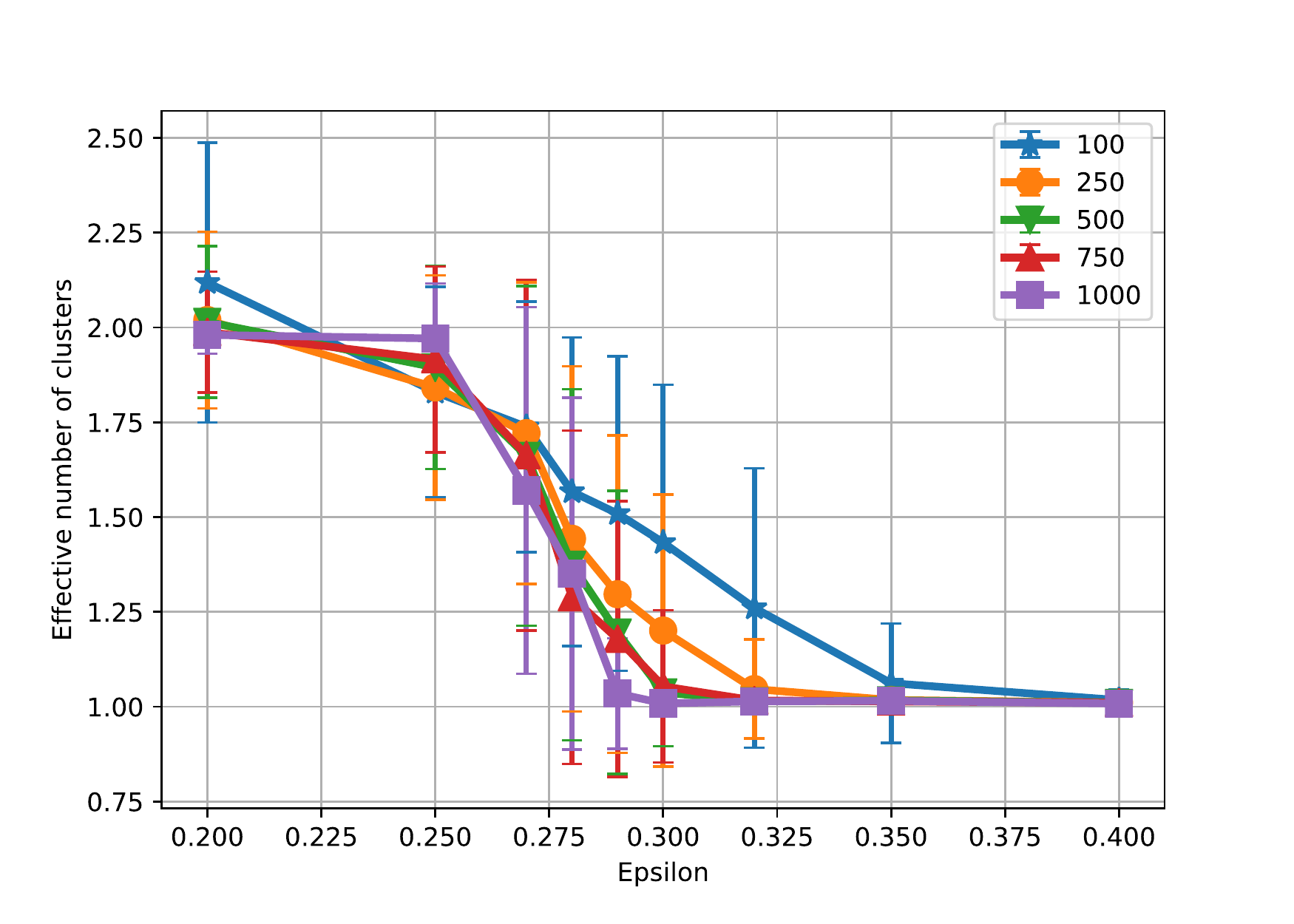}
\caption{{\bf Number of clusters without algorithmic bias.} Effective number of clusters obtained for various $\varepsilon$ and $N$, when $\gamma=0$.  \alina{Values are averaged over 200, 125, 85, 45 and 45 runs for $N \in \{100,250,500,750,1000\}$, respectively. Error bars show one standard deviation from the mean.}}  
\label{fig_finite_size_deffuant}
\end{figure}

A third analysis that we performed aimed at understanding whether the size of the population plays a role in the  effect of the algorithmic bias. Again, this is important for realistic scenarios, since opinion formation may happen both at small and at large scale. Hence we look at the transition between consensus and segregation for variable population sizes, both for the original and for the extended model.

\begin{figure}
\centering
\includegraphics[width=0.7\columnwidth]{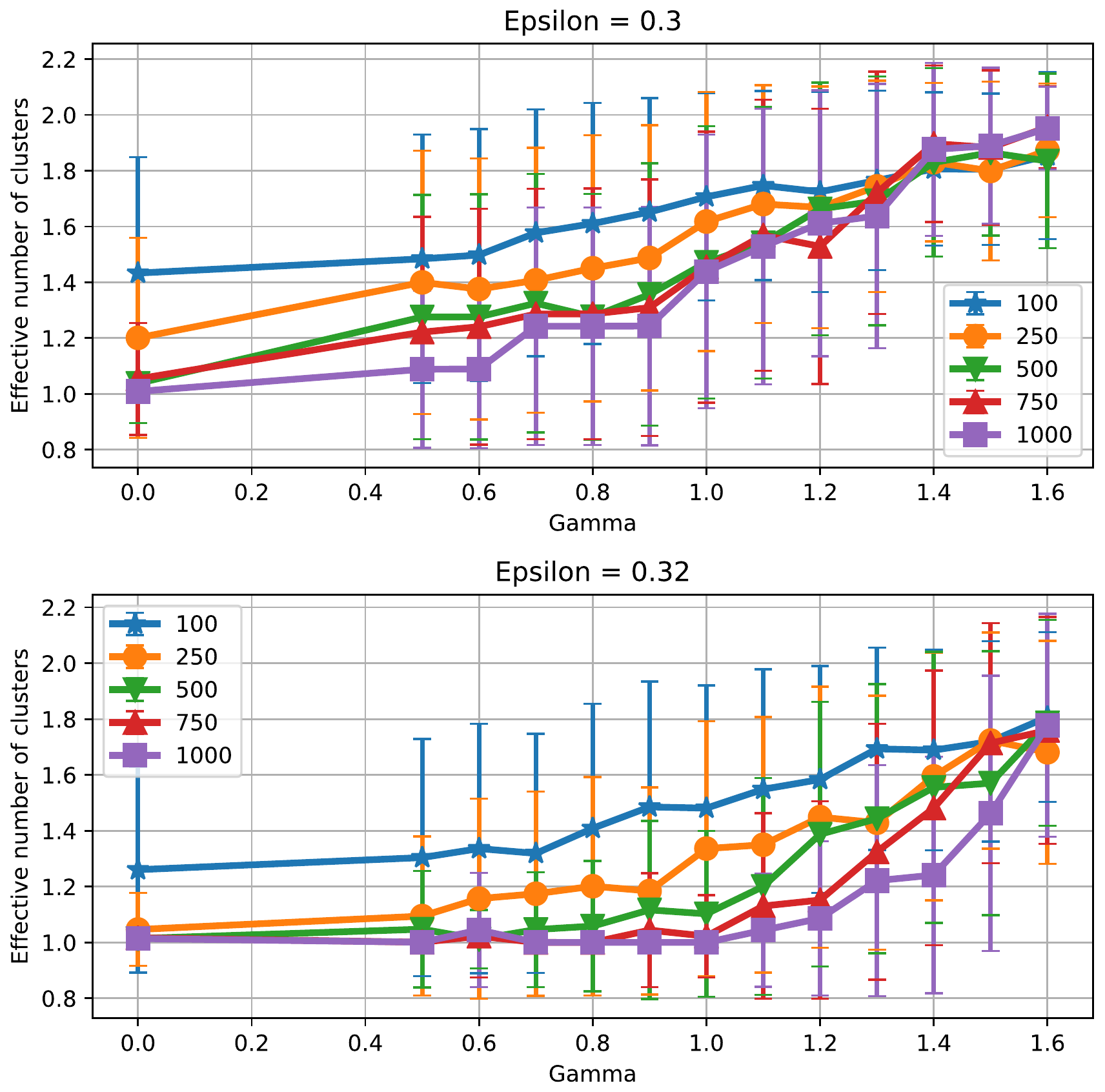}
\caption{{\bf Number of clusters with algorithmic bias.} Effective number of clusters obtained for  various $\varepsilon$, $\gamma$ and $N$. \alina{ Values are averaged over 200, 125, 85, 45 and 45 runs for $N \in \{100,250,500,750,1000\}$, respectively. Error bars show one standard deviation from the mean.} }
\label{fig_finite_size_gamma}
\end{figure}

In the original model, the transition between $c$ and $c+1$ clusters is shown to be continuous, with an interval for $\varepsilon$ where both cases can appear in different simulation instances (see Fig~4 in~\cite{Deffuant2000}). In particular, the transition between one and two clusters happens for $\epsilon \in (0.25,0.3)$. We performed numerical simulations to test whether this interval changes with $N$, given that, to the authors' knowledge, a detailed study in this direction does not exist for the bounded confidence model. 
Fig~\ref{fig_finite_size_deffuant} shows the \alina{mean} effective number of clusters over multiple runs obtained with $N\in\{100,250,500,750,1000\}$. \alina{Error bars represent one standard deviation from the mean.} It is clear that, as $N$ increases, the transition becomes more abrupt, i.e. the transition interval decreases in length, and gets closer to $\varepsilon=0.25$.
For very small populations, in particular, \alina{the number of clusters in the transition area is larger, probably due to  a lower density of opinions in the starting population, which facilitates formation of clusters even when larger populations would converge to consensus. Hence we can conclude that a small $N$ can also favour segregation. }  \alina{The error bars are almost invisible outside the phase transition interval, but quite large inside this interval. This is due to the fact that within the transition interval some simulations converge to one cluster, while other to two clusters, obtaining thus large standard deviations from the mean.}
 analyse the effect of the population size when $\gamma >0$, we consider two different $\varepsilon$ values (0.3 and 0.32). These were chosen because for $\gamma=0$ they yield one cluster, while segregation emerges as $\gamma$ grows. Fig~\ref{fig_finite_size_gamma} shows the effective number of clusters obtained for various population sizes. Again, the transition from one to two clusters is more steep in $\gamma$ as $N$ increases, with small population sizes favouring segregation. In these conditions, it seems that algorithmic bias can actually be more efficient in hindering consensus for smaller groups.

\subsection{Effect of the initial condition}\label{sec:init}

\begin{figure}
\centering
\includegraphics[width=0.7\columnwidth]{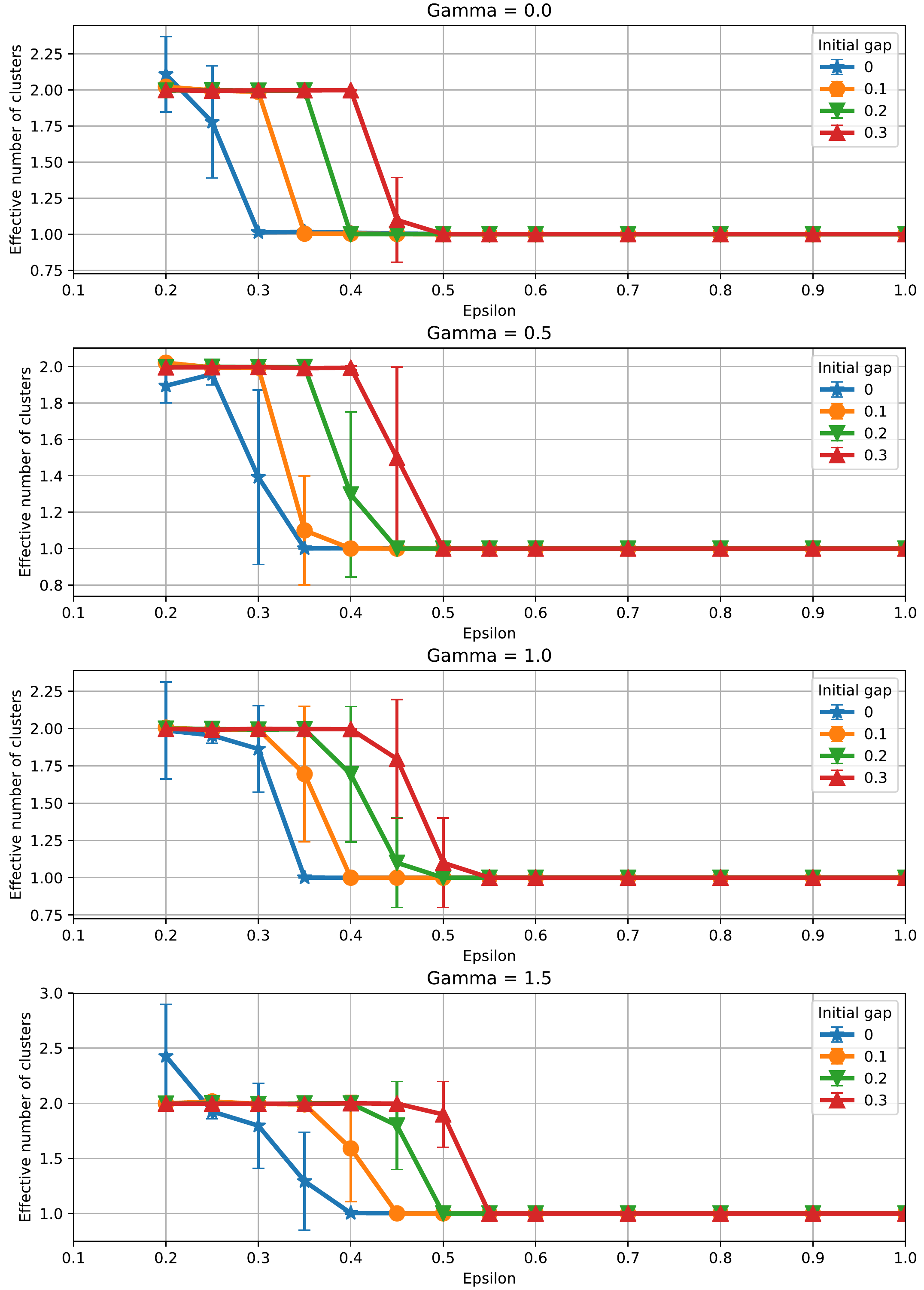}
\caption{{\bf Initial condition.} Effect of the initial condition on the effective number of clusters (averages over 10 runs).   }
\label{fig_gap_clusters}
\end{figure}

\begin{figure}
\centering
\includegraphics[width=0.7\columnwidth]{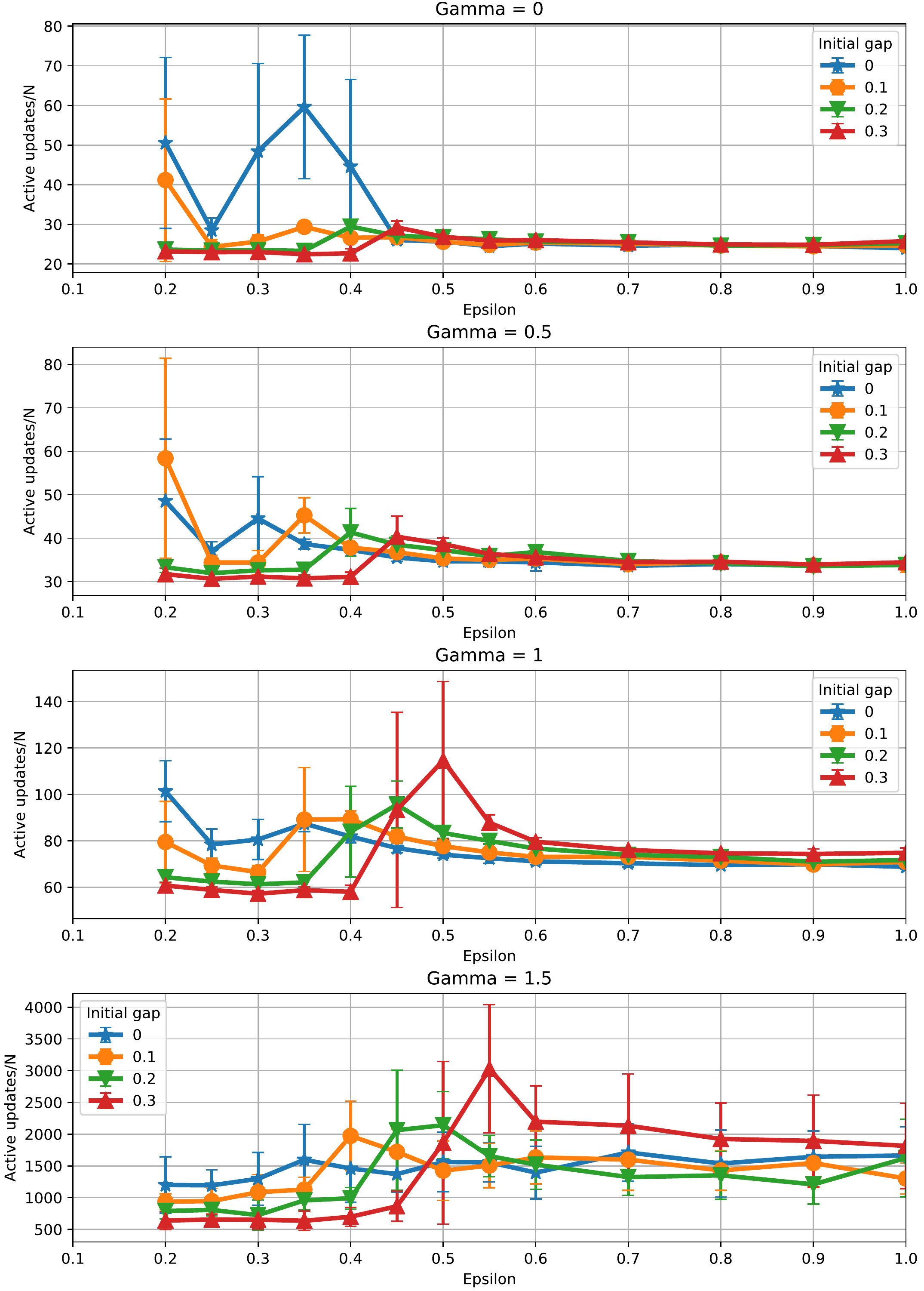}
\caption{Effect of the initial condition on the convergence time measured in number of active interactions (averages over 10 runs).  }
\label{fig_gap_time}
\end{figure}

Previous results were obtained for the case where numerical simulations assumed uniformly random initial opinions in the population. However, in reality, opinion formation may start from slightly fragmented initial conditions. To simulate this, we introduced artificially a symmetric gap around the opinion value $0.5$, to simulate a population where opinions are already forming. The width of the gap was varied to understand the effect of various fragmentation levels both for the original bounded confidence model ($\gamma=0$) and for our extension.

Fig~\ref{fig_gap_clusters}  shows the \alina{mean} effective number of clusters obtained for various gap sizes, \alina{with error bars showing one standard deviation from the mean}. For the original model, the gap shifts the transition from two to one cluster towards larger values of $\epsilon$. Hence a fragmented initial condition favors fragmentation even later during the evolution of opinions. However, as $\epsilon$ grows, fragmentation disappears, hence a higher tolerance level 
can overcome a fragmented initial condition. We also note that as $\gamma$ grows, the two effects from algorithmic bias and initial fragmentation add up to push the transition to consensus ever closer to $\epsilon=1$.  \alina{Error bars show again how in the transition interval the population converges to one cluster is some simulations and to two clusters in others, resulting in relatively large deviations from the mean. However, outside the transition interval error bars are almost invisible, hence the number of cluster is very stable from one simulation to another.}

In terms of time for convergence, Fig~\ref{fig_gap_time} plots the number of \emph{active interactions} for the case of a fragmented initial condition. It appears that  initial fragmentation speeds up convergence when the final population is also fragmented. However, when the final population reaches consensus (one cluster), the effect is reversed, i.e. initial fragmentation slows down convergence. The time to convergence continues to grow very fast with $\gamma$, as seen previously for a uniform initial condition.

Hence, again, the two effects appear to work together against reaching consensus, either by favoring the appearance of additional clusters or by slowing down consensus when this could, in principle, emerge.

\section{Discussion and Conclusions}

A model of algorithmic bias in the framework of bounded confidence was presented, and its behavior analyzed. Algorithmic bias is a mechanism that encourages interaction among like-minded individuals, similar to patterns observed in real social network data. We found that, for this model,  algorithmic bias hinders consensus and favors opinion segregation through two different mechanisms. On one hand, consensus is hindered by a very strong slowdown of convergence, so that even when one cluster is asymptotically
obtained, the time to  reach it is so long that in practice consensus will never appear. Additionally, we observed segregation of the population as the bias grows stronger, with the number of clusters obtained increasing compared to the original model. A fragmented initial condition also enhances the fragmentation, augmenting the effect of the algorithmic bias. Additionally, we observed that small populations may be less resilient to segregation, due to finite size effects.

The results presented here are based on the assumption that bounded confidence exists, i.e. individuals with very distinct opinions do not exchange information hence do not influence each other. However, our conclusions regarding the fact that algorithmic bias hinders consensus still stand even when  bounded confidence is removed (i.e. $\varepsilon=1$). In this case, consensus still becomes extremely slow as the bias increases, hence is never achieved in practice, a result that we believe will apply to any other model of opinion dynamics. It would also be interesting to see how taking into account a more realistic social network structure among individuals, instead of a complete graph where anybody may interact with anybody else, would impact the opinion formation process, possibly exacerbating the effects observed in this study.

Although there is evidence that many types of social interactions are subject to algorithmic bias, the debate still continues on whether this generates or not opinion segregation in the long term. Our numerical results support the first option, which we plan to analyse in more detail in the future by applying our model to real data from social network processes. We would also like to understand how external information  could affect the behavior observed, especially when the sources of information are also selected based on a similar bias. \alina{Recent work on how to counteract opinion polarisation on social networks has also appeared~\cite{Garimella2017}, and initial results suggest that facilitating interaction among chosen individuals in polarised communities can alleviate the issue. We will also investigate this with our model.  }

\section*{Acknowledgements}
We thank 
the IT Center of the University of Pisa (Centro Interdipartimentale di Servizi e Ricerca) for providing access to computing resources for simulations. This work was supported by   the European Community's H2020
Program under the funding scheme ``FETPROACT-1-2014:
Global Systems Science (GSS)", grant agreement \# 641191
CIMPLEX ``Bringing CItizens, Models and Data together
in Participatory, Interactive SociaL EXploratories".

\bibliographystyle{abbrev}

\end{document}